\documentclass[seceq,addenda]{ptptex}
\usepackage[xdvi,dvips]{graphicx}

\title{% 
Chiral and Color-superconducting 
Phase Transitions with Vector Interaction \\
in a Simple Model
}

\author{%       %Use \scshape  for the family name
Masakiyo \textsc{Kitazawa}, 
Tomoi \textsc{Koide}, 
Teiji \textsc{Kunihiro} and
Yukio \textsc{Nemoto}
}

\publishedin{
Prog. Theor. Phys. Vol.108 (2002), 929}

\begin{document}

\maketitle

In the above paper, we have shown that the critical line of the
 first order chiral transition of QCD can have
 {\em two} endpoints.
In this addendum,  we elucidate the mechanism
to realize the two-endpoint structure in the QCD phase diagram
and argue the robustness for the appearance of
such an interesting phase structure.
\\

Recently, there is growing interest in the existence and its
phenomenological consequences of  the {\em endpoint} of 
the first order chiral phase transition
(or a tricritical point in the chiral limit) of QCD
at finite temperature ($T$) and density.
In the above paper, we have demonstrated that 
there can appear {\em another} endpoint of the critical line of the
first order transition  in the lower $T$ region
 where the color superconductivity (CS) comes into play;
we shall
 call such a phase structure in the phase diagram the two-endpoint
structure (abbreviated to TEPS).
After the publication of our paper, some colleagues asked us in what condition
such another end point in the lower $T$ region can appear.
Indeed, we are aware that 
 we  have failed in given sufficient discussions on  
TEPS in the paper.
In this addendum,  partly to answer the questions asked by the colleagues,
we shall try to clarify the mechanism
for TEPS to be realized in  a qualitative way
and show that the growth of the gap $\Delta$ of CS,
which is caused by the larger vector coupling $G_V$ in our case,
is responsible for the realization of TEPS.

To understand the mechanism 
of the appearance of another endpoint in the lower $T$ region,
it is best to begin with recalling the way 
how the endpoint of the chiral transition in the {\rm higher}
 $T$ side gets to exist.
At finite $T$,
 there exist quarks with a high momentum well above the
Fermi momentum as described by the Fermi-Dirac (F-D) distribution function
 $n(p) = 1/(e^{(p-\mu)/T}+1)$; see  Fig.~1(a).
An important point is that
 the positive energy states of quarks contribute 
positively to the quark condensate
$\langle\bar{q}q\rangle \propto M_D$, making the net value of it smaller.
The  condensate is expected to vary 
 smoothly with the change of  $T$ as given  by the F-D distribution function.
Thus the chiral phase transition becomes weaker and 
the order of it changes from a first to crossover
(or second  in the chiral limit) when $T$ is raised, as it is. 

Now how does the diquark condensate or the gap $\Delta$ of CS
affect the chiral transition?
Here it is worth mentioning, although well known again, that CS 
with $\Delta$ leads to a quark distribution function similar to 
the F-D distribution function $T\not=0$:
The distribution function for the massless quarks 
with $\Delta$ at $T=0$ reads
\begin{eqnarray}
n(p) = \frac12
\left( 1 - \frac{p-\mu}{ \sqrt{ (p-\mu)^2+\Delta^2 } } \right),
\end{eqnarray}
which behaves as shown in Fig.~1(b); notice the similarity of Fig.~1(b) with
Fig.~1(a).
Thus one can naturally expect that the larger $\Delta$ also weakens
the chiral phase transition at low $T$.
To see the similarity of the role of $\Delta $ and $T$
 on the chiral condensate more transparently, let us 
rewrite  the gap equation Eq.~(3.13) in the text as follows:
\begin{eqnarray}
M_D &=& 8 G_S M \int \frac{ d^3 p }{ (2\pi)^3 }
\frac1{ E_p } \left\{ 1 - n_N^- ( p ) - n_N^+ ( p )
+ 2 \left( 1 - n_S^-( p ) - n_S^+( p ) \right) \right\},
\label{eqn:GEQ}
\end{eqnarray}
where
$n_S^\pm = (1/2) ( 1 - \xi_\pm/\epsilon_\pm \cdot\tanh(\beta\epsilon_\pm/2))$
denotes the distribution functions of the color superconducting
quarks and anti-quarks,
and $n_N^\pm = 1/(e^{(p-\mu)/T}+1)$ corresponds to the normal ones.
Notice that Eq. (\ref{eqn:GEQ}) depends on $\Delta$ and $T$
only through $n_N^\pm$ and $n_S^\pm$, which makes
 the effect of $\Delta$ on the chiral condensate
is hardly distinguishable with that of $T$.
It implies that  if a sufficiently large gap is formed near 
the chiral transition,
 the phase transition can change from a  first order 
to crossover.

\begin{figure}[t]
\begin{center}
{\includegraphics[scale=1.1]{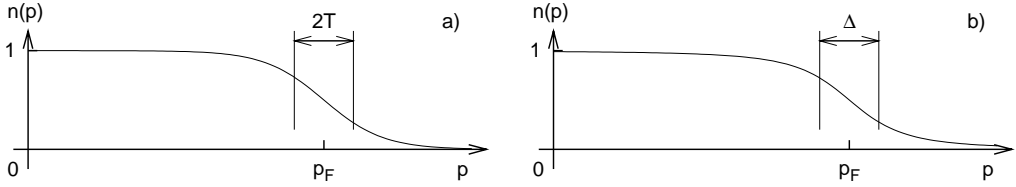}}
\caption{
The typical examples of the distribution functions for
(a) finite temperature , $n = 1/(e^{(p-\mu)/T}+1)$
and (b) finite diquark gap $\Delta$ Eq.~(0.2).}
\end{center}
\end{figure}

What is  the role of the  vector coupling $G_V$ in this problem?
As is shown in the paper,
the larger $G_V$  not only makes the first order chiral transition weaker
in the higher $T$ region but also  widens the region in the phase diagram
of the coexisting phase where $M_D$ and $\Delta$ are  both finite. 
Thus a larger $\Delta$ can be realized with $G_V$ 
around the critical line of the chiral transition in the lower $T$ region, 
which in  turn leads to the weakening of  the phase transition 
owing to the above mentioned mechanism.
Since lower the $T$, larger the $\Delta$,
the weakening of the phase transition is more effective
at lower $T$; thus
  the first order transition in the low $T$ region
starts to change into a crossover one  
from $T=0$ with some $G_V$.
This is how another endpoint gets to exist at
 relatively lower $T$; hence realized is TEPS.
As $G_V$ is increased further,
 the lower endpoint approaches toward higher $T$,
and eventually
the crossover  transition at lower $T$ merges with that extending 
from the higher $T$ region;
the first order transition ceases to exist
in the phase diagram.

It should be emphasized that
TEPS does not appear without incorporating
the quark-quark (qq) interaction
which causes the interplay between $\chi$SB and CS.
The above mechanism for the appearance of TEPS
 is considered independent of models employed,
provided that the interplay between the $\chi$SB and CS 
is taken into account.

Although TEPS is obtained
in an effective model incorporating $G_V$ in the present case,
it is worth mentioning that $G_V$ is actually  dispensable
for the manifestation of TEPS:
When the chiral transition is first order but already sufficiently weak
without the qq interaction,  the incorporation of the 
qq interaction acts to realize
TEPS by further weakening the first order transition at lower $T$.
Therefore, the second end point could  get to exist, for example, 
if the scalar  interaction as given by $G_S(\bar{q}q)^2$
which is the driving force of the $\chi$SB were  much weaker.

Finally, we remark that TEPS discussed  here
and the mechanism for the appearance of it
are completely different from the ones given in \citen{ref:TK},
which discusses 
possible effects of {\em isospin} chemical potential 
on the chiral transition.


\begin{thebibliography}{99}
\bibitem{ref:TK}
D.~Toublan, and J.B.~Kogut, Phys. Lett. B564 (2003) 212
(arXiv:hep-ph/0301183).
\end{thebibliography}
\end{document}